\documentstyle[12pt, aasms4, epsfig]{article}
\catcode`@=11
\def\eqalign#1{\null\,\vcenter{\openup\jot\m@th
  \ialign{\strut\hfil$\displaystyle{##}$&$\displaystyle{{}##}$\hfil
      \crcr#1\crcr}}\,}
\def\eqalignleft#1{\null\,\vcenter{\openup\jot\m@th
  \ialign{\strut$\displaystyle{##}$\hfil&$\displaystyle{{}##}$\hfil
      \crcr#1\crcr}}\,}
\catcode`@=12

\def\lax    {\ifmmode{_<\atop^{\sim}}\else{${_<\atop^{\sim}}$}\fi}
\def\gax    {\ifmmode{_>\atop^{\sim}}\else{${_>\atop^{\sim}}$}\fi}
\def\kms    {\ifmmode{{\rm ~km~s}^{-1}}\else{~km~s$^{-1}$}\fi}

\def\bk{\lower 6pt\hbox{${\buildrel k\over \sim}$}}
\def\bv{\lower 6pt\hbox{${\buildrel v\over \sim}$}}

\begin{document}
\doublespace
\textwidth 6.7truein
\textheight 9.5truein
\topmargin -1cm
\hoffset = 0.5truein
\title{\bf FORMATION AND EVOLUTION OF THE TRANS-NEPTUNIAN BELT 
AND DUST}
\author{\bf Sergei Ipatov\altaffilmark{1}}
\affil{M.V. Keldysh Institute of Applied Mathematics, RAS, Moscow;
ipatov@keldysh.ru}
\author{\bf Leonid M. Ozernoy\altaffilmark{2}}
\affil{5C3, School of Computational Sciences and Department of Physics
\& Astronomy,\\ George Mason U., Fairfax, VA 22030-4444; also Laboratory for
Astronomy and Solar\\ Physics, NASA/Goddard Space Flight Center, Greenbelt,
MD 20771}
\altaffiltext{2}{Corresponding author. Fax: $+1$-301-286-1617; e-mail:
ozernoy@science.gmu.edu, ozernoy@stars.gsfc.nasa.gov; 
{\tt http://science.gmu.edu/$^\sim$ozernoy}}
\bigskip
\centerline{\bf Abstract}

  Trans-Neptunian objects (TNOs) with diameter greater than 100 km currently 
moving in not too eccentric orbits could be formed directly by the contraction 
of large rarefied condensations. Along  with the gravitational influence of 
planets, gravitational interactions of TNOs played a certain role in their 
orbital evolution as well. More than 20\% of Earth-crossing objects could have 
come from the trans-Neptunian belt. TNOs and Centaurs (invisible comets mainly 
beyond Jupiter) could produce an important contribution to the dust content of 
the interplanetary dust cloud.

\medskip

Paper submitted to the International Conference "Kazan 
Astronomy 2001" (September 24-29, 2001, Kazan, Russia).

\newpage
\section*{Introduction}

   So far more than 400 trans-Neptunian objects (TNOs) are known. Jewitt et al. 
(1996) estimated the total mass of the present Edgeworth-Kuiper belt (EKB) for 
objects with $30 \le a \le 50$ AU to be $0.06m_\oplus$ to $0.25m_\oplus$, where 
$m_\oplus$ is the mass of the Earth. For TNOs with $a \le 50$ AU, the average 
values of eccentricity and inclination are evaluated to be $e_{av} \approx 0.1 $ 
and $i_{av} \approx 8^o$, respectively. Objects moving in highly eccentric 
orbits (mainly with $a>50$ AU) are called "scattered disk objects" (SDOs). For 
SDOs $e_{av} \approx 0.5 $ and  $i_{av} \approx 16^o$. The total mass of SDOs in 
eccentric orbits between 40 and 200 AU has been estimated by different authors 
in the range $(0.05-0.5)m_\oplus$. According to Duncan et al. (1995), about 10-
20\% TNOs with $a<50$ AU left the EKB during the last 4 Gyr under the 
gravitational influence of planets, and about 1/3 of Neptune-crossing objects 
could reach Jupiter's orbit during their lifetimes. For a detailed review of the 
formation and migration of celestial bodies in the Solar system, see Ipatov 
(2000).

   The sources of the interplanetary dust (IPD) cloud cannot be entirely reduced 
to comets and asteroids alone. Several factors indicate that the overall dust 
production rate from TNOs may not be negligible compared to that of comets and 
hence a third important component of the IPD cloud (besides interstellar grains) 
might be the EKB, or `kuiperoidal' dust. In our opinion, the EKB influences the 
formation of the IPD cloud in two ways: (i) as a source of small-size particles 
slowly drifting toward the Sun under a combined action of the 
Poynting--Robertson drag and perturbations from the planets; and (ii) as a 
source of millions of comets between Jupiter and Neptune (Levison \& Duncan, 
1997; Ozernoy, Gorkavyi, \& Taidakova, 2000a), which, in turn, serve as 
additional sources of dust. The dust can be produced due to evaporation of the 
volatile material from the TNO surface as a result of a variety of processes, 
such as the Solar wind and the heating by the Sun, micrometeor bombardment, 
mutual collisions of kuiperoids, etc. 

\section*{Formation of the trans-Neptunian belt}

    Stern (1996), Stern and Colwell (1997), Davis and Farinella (1997), and 
Kenyon and Luu (1999), among others, investigated formation and collisional 
evolution of the EKB. In their models, the process of accumulation of TNOs in 
the massive EKB took place at small (usually about 0.001) eccentricities. Ipatov 
(2000) found that, due to the gravitational influence of the forming giant 
planets and the mutual gravitational influence of planetesimals, such small 
eccentricities could not exist during all the time needed for the accumulation 
of TNOs. In Ipatov's runs the maximal eccentricities of TNOs always exceed 0.05 
during 20 Myr due to the gravitational influence of the giant planets. Eneev 
(1980) hypothesised that large TNOs and the planets were formed by the 
accumulation of large rarefied dust-gas condensations, which later contracted to 
a solid body density. We do not think that all the planets could form in such a 
way; however, TNOs with diameter $d\ge100$ km could be formed mainly by the 
contraction of large rarefied dust condensations, and not by the accretion of 
smaller solid planetesimals. Perhaps, the largest asteroids and planetesimals 
with $d\ge100$ km in the zone of the giant planets could be formed in the same 
way, while a part of smaller objects could be mainly debris of larger objects 
and another part could formed directly by contraction of condensations. Even if 
the sizes of initial condensations, which had been formed, due to gravitational 
instability, from the circumsolar dust disk, were about the same at some time at 
the same distance from the Sun, it would be, due to mutual interactions, a 
distribution in masses of the final condensations, which then contracted into 
planetesimals. As in the case of accumulation of planetesimals, there could be a 
"run-away" accretion of condensations. Possibly, during the time needed for 
contraction of condensations into planetesimals, some largest final 
condensations could reach such masses that they formed planetesimals as large as 
several hundred kilometers. 

    During accumulation of the giant planets, planetesimals of large 
eccentricities with the total mass of several tens $m_\oplus$ could move from 
the feeding zone of the giant planets to the trans-Neptunian region (Ipatov, 
1987, 1993). These planetesimals increased eccentricities of orbits of 'local' 
TNOs, which initial mass could exceed $10m_\oplus$, and swept most of them. A 
small portion of such planetesimals could survive in eccentrical orbits beyond 
Neptune's orbit and become SDOs. The total mass of the planetesimals that came 
from behind Jupiter's orbit during formation of the giant planets and then 
collided the Earth is about the mass of water in the Earth's oceans (Ipatov, 
1995). Dynamical lifetimes of most planetesimals located inside Neptune's orbit 
were less than 100 Myr, while those for the objects beyond Neptune were much 
larger. Therefore, the objects that came from eccentric orbits located mainly 
beyond Neptune's orbit, were dominating at the end of an intense bombardment  of  
terrestrial planets, which  finished 4 Gyr ago.

\section*{Collisional  evolution of the trans-Neptunian belt}

    Frequency of collisions of bodies in the EKB and in main asteroid belt (MAB) 
was evaluated by Stern (1996), Davis and Farinella (1997), Durda and Stern 
(2000), and Ipatov (1995, 2000). It is assumed that there are about 106 
asteroids with $d\ge 1$ km in the MAB, and the number of asteroids with $d>D_*$ 
is proportional to $D_*^{-\alpha}$, with $\alpha$ between 2 and 2.5 (Binzel {\it 
et al.}, 1991). In the MAB for the ratio $s$ of masses of two colliding bodies, 
for which a collisional destruction of a larger body takes place, equal to 
$10^4$, a collisional lifetime $T_c$ of a body with $d=1$ km is about 1 Gyr 
(Ipatov, 1995). If $\alpha=2$ and $s$=const, then a value of $T_c$ does not 
depend on $d$. It is considered that for near-Earth objects (NEOs) $\alpha>3$ at 
$d<40$ m. If it so also for the MAB, then $T_c<2$ Myr for 1-m rocky asteroid 
(Ipatov, 1995). At $\alpha=2$, $s=10^3$, and the mass of the EKB $\sim 
0.1m_\oplus$ for TNOs with $d\ge100$ km, one gets $T_c=30$ Gyr. For $s=10^4$ 
(and $\alpha=2$) the values of $T_c$ are smaller by a factor of 4.6 than those 
for $s=10^3$. An 1-km TNO collides with one of $10^{12}$ 100-m objects on 
average once in 3 Gyr. Therefore, at $s$=const, the values of $T_c$ for 1-km 
TNOs are of the same order of magnitude as those for main-belt asteroids. 

   The total mass of SDOs moving in highly eccentric orbits between 40 and 200 
AU is considered to be of the same order or greater than the total mass of the 
EKB, and the mean energy of a collision of a SDO with a TNO is greater 
(probably, by a factor of 4) than that for two colliding TNOs of the same 
masses. Therefore, though SDOs spend a smaller part of their lifetimes at 
distances $R<50$ AU, the probability of destruction of a TNO with $30<a<50$ AU 
by SDOs can be of the same order of magnitude (possibly, even larger) than that 
by TNOs. 

    Ipatov (1995, 2000) showed that during the last 4 Gyr several persents of 
TNOs could change their semimajor axes by more than 1 AU due to the 
gravitational interactions with other TNOs. First estimates of gravitational 
interactions between TNOs were made by Ipatov long before the first TNO was 
found in 1992. Even small variations in orbital elements of TNOs due to their 
gravitational influence and collisions could cause large variations in orbital 
elements of TNOs under the gravitational influence of planets (Ipatov and 
Henrard, 2000).

\section*{Migration of bodies to the Earth}

    The orbital evolution of a hundred TNOs under the gravitational influence of 
planets is described by Ipatov (1999, 2000) and Ipatov and Henrard (2000). 
During the evolution, the perihelia of orbits of two test objects decreased by 1 
AU during 25 and 64 Myr, respectively. Numerical integration of the average time 
interval, during which an object crosses Jupiter's orbit during its lifetime, is 
0.2 Myr; the fraction of Jupiter-crossing objects (JCOs), which reach the 
Earth's orbit during their lifetimes amounts to 0.2; and the average time, 
during which an JCO crosses the orbit of Earth, is about 5000 yr. Using these 
results, Ipatov (1999, 2000) found that if the number of 1-km EKBOs is as large 
as $10^{10}$  (Jewit et al., 1996), then the number of present JCOs of $d\ge1$ 
km that came from the trans-Neptunian belt amounts to $3\cdot 10^4$, and about 
170 former TNOs cross both the orbits of Earth and Jupiter. These objects 
represent about 20\% of ECOs, if the number of 1-km Earth-crossing objects is 
750. A lot of former TNOs can move in Encke-type orbits with aphelia inside 
Jupiter's orbit. If nongravitational forces are included into interactions  (in 
impulse approximation), the rate at which objects could be decoupled from 
Jupiter and attain orbits like those of NEOs is increased by a factor of four or 
five (Asher et al., 2001).

\section*{Modeling of the interplanetary dust}

   Until recently, the main stumbling block to implementing the comprehensive 
study of IPD has been the absence of a physical model for the IPD cloud. Such a 
model would establish a link between the observable characteristics of the 
zodiacal cloud and the dynamical and physical properties of the parent minor 
bodies of the Solar system. A preliminary physical model of the IPD cloud based 
on a new computational approach elaborated by Gorkavyi, Ozernoy, Mather, \& 
Taidakova is described in detail by Ozernoy (2001). This approach permits with 
modest computational resources to integrate trajectories of hundreds of 
particles and to effectively store up to $10^{10}-10^{11}$ particle positions as 
if they were real particles, which provides a high fidelity 3D distribution of 
the dust. An appreciable increase in statistics, compared to e.g. Liou \& Zook 
(1999), brings a factor of $10^4$ improvement in the detail of a model and 
enables us to model the IPD cloud at a qualitatively new, 3-D level. Moreover, 
our approach makes it possible to study, besides stationary processes, certain 
non-stationary processes as well, e.g. evolution toward steady-state 
distributions, dust production from non-steady sources, decrease in particle 
size (due to evaporation and sputtering) and number (due to collisions), etc.

   The numerical codes employed account for the major dynamical effects that 
govern the motion of IPD particles: the Poynting--Robertson drag and solar wind 
drag; the solar radiation pressure; particle evaporation; gravitational 
scattering by the planets; and the influence of mean-motion resonances.

\section*{The simulated distribution of kuiperoidal dust}

   The efficiency and power of the employed codes mentioned in the previous 
section, has been demonstrated by performing the following simulations: (i) 
distribution of the scattered comets, which enables one to reveal the four 
`cometary belts' associated with the orbits of four giant planets  (Ozernoy et 
al., 2000a), which are expected to contain 20-30 million of cold comets; (ii) 
detailed analysis of a rich resonant structure found in these belts, which 
predicts the existence of gaps similar to the Kirkwood gaps; (iii) a 3-D 
physical model of the IPD cloud, which explains the available data of Pioneers 
and Voyagers dust detectors; (iv) zodiacal light distribution in the Solar 
system, which fits the COBE data with an average accuracy of 0.85\%, and (v) 
resonant structure in dusty circumstellar disks of Vega and Epsilon Eridani 
(Ozernoy et al., 2000b) and a warp in dusty disk of Beta Pictoris considered to 
be  a signature of embedded extrasolar planets.

       Under a set of reasonable assumptions, it seems safe to conclude: 

  1. The kuiperoidal dust plays a role more important than previously 
recognized. It appears to account for the space dust observations beyond 6 AU, 
while near Earth it could possibly contribute as much as 1/3 of total number 
density (1/4 of surface density) and 1/3 of the zodiacal emission near ecliptic.
 
  2. The two other components of the IPD cloud, the cometary and asteroidal dust 
contribute respectively 36\% and 30\% of the number density and the zodiacal 
emission (at ecliptic) near Earth. The cometary particles contribute 60\% to the 
surface density of the IPD cloud  near Earth. A solely two-component model (i.e. 
without the kuiperoidal dust) would give a worse fit of dust distribution at 
Earth and would fail entirely for the outer Solar system.

   3. Further improvements in the IPD modeling will include, among others, 
particles of different sizes, account for evaporation and sputtering of dust as 
a function of heliocentric distance, and include short-term (days to months) 
variability and small-scale phenomena in the zodiacal cloud.
                                                              
\section*{Acknowledgements}

   We acknowledge support of this work by NASA grant NAG5-10776, the Russian 
Federal Program "Astronomy" (section 1.9.4.1), Russian Foundation for Basic 
Research (01-02-17540), and INTAS (00-240).


\section*{References}


\def\ref#1  {\noindent \hangindent=24.0pt \hangafter=1 {#1} \par}
\smallskip

    \ref{Asher, D.J., Bailey, M.E., and Steel, D.I., 2001, In "Collisional 
processes in the solar system". Ed. by M. Ya. Marov and H. Rickman, ASSL.}
    \ref{Binzel, R.P., Barucci, M.A., and Fulchignoni, M, 1991, Sci. American, 
265, October, 66-72.}
    \ref{Jewitt, D., Luu., J. and Chen, J., 1996, Astron. J., 112, 1225-1238.}
    \ref{Davis, D. R. and P. Farinella, 1997, Icarus, 125, 50-60.}
    \ref{Duncan, M.J., Levison, H.F. and Budd, S.M., 1995, Astron. J., 110, 
3073-3081.}
    \ref{Durda, D.D. and Stern, S.A., 2000, Icarus, 145, 220-229.}
    \ref{Eneev, T.M., 1980, Sov. Astron. Letters, 6, p. 295-300 in Russian 
edition.}
    \ref{Ipatov, S.I., 1987, Earth, Moon, and Planets, 39, 101-128.}
    \ref{Ipatov, S.I., 1993, Solar System Research, 27, 65-79.}
    \ref{Ipatov, S.I., 1995, Solar System Research, 29, 261-286.}
    \ref{Ipatov, S.I., 1999, Celest. Mech. Dyn. Astron., 73, 107-116.}
    \ref{Ipatov, S.I., 2000, "Migration of celestial bodies in the Solar 
System", Editorial URSS, Moscow, (in Russian), 320pp.}
    \ref{Ipatov, S.I. and Henrard, J., 2000, Solar System Research, 34, 61-74.}
    \ref{Kenyon, S. J., and Luu, J. X., 1999, Astron. J., 118, 1101-1119.}
    \ref{Levison, H.F. and Duncan, M.J., 1997, Icarus, 127, 13-23.}
\ref{Liou, J.-C. and Zook, H.A., 1999, Astron. J., 118, 580.}
\ref{Levison, H.F. and Duncan M.J. ,1997, Icarus, 127, 13-23}
\ref{Liou, J.-C. and Zook, H.A., 1999, Astron. J., 118, 580}
 \ref{Ozernoy, L.M., Gorkavyi, N.N., and Taidakova, T., 2000a,  
Planetary Space Science, 48, 993}
   \ref{Ozernoy, L.M., Gorkavyi, N.N., Mather, J.C. and Taidakova, T., 2000b, 
Astrophys. J., 537, L14.}
   \ref{Ozernoy, L.M., 2001, in "The Extragalactic Infrared Background and its
    Cosmological Implications" (IAU Symp. No.204).Eds. M. Harwit and
    M.G. Hauser. ASP Conference Series, p. 17.}
    \ref{Stern, S. A., 1996, Astron. J., 112, 1203-1211.}
    \ref{Stern, S. A. and Colwell, J. E., 1997, Astron. J., 114, 841-849.}

\end{document}